\documentclass[showpacs,aps,pre,twocolumn]{revtex4}

\usepackage{graphicx}
\usepackage{fncylab,hyperref,environ}
\usepackage{dcolumn}
\usepackage{url}

\usepackage{amssymb}
\usepackage{amsmath}

\usepackage{tikz}
\usepackage{pgfplots}
\usepackage{pgfplotstable}

\newcommand{\trc}{\operatorname{tr}}

\labelformat{figure}{Figure #1}
\labelformat{equation}{(#1)}
\labelformat{section}{Section #1}
\labelformat{subsection}{.#1}
\labelformat{table}{Table #1}

\newcommand{\myinsetwidth}{0.22\textwidth}

\pgfplotscreateplotcyclelist{list1}{
    {color=red, mark=triangle, only marks},
    {color=brown, mark=diamond, only marks},
    {color=violet, mark=square, only marks},
    {color=green!60!black, mark=o, only marks},
    {color=blue, mark=pentagon, only marks},
    {color=gray, mark=triangle*, mark size=1.3, only marks},
    {color=magenta,mark=diamond*, mark size=1.3, only marks},
    {color=orange, mark=square*, mark size=0.8, only marks},
    {color=teal, mark=otimes*, mark size=0.9, only marks},
    {color=olive, mark=pentagon*, mark size=1.3, only marks}}

\pgfplotscreateplotcyclelist{list2}{
    {color=red},
    {color=brown},
    {color=violet},
    {color=green!60!black},
    {color=blue},
    {color=gray},
    {color=magenta},
    {color=orange},
    {color=teal,},
    {color=olive}}

\makeatletter

\begin{document}

\title{Jammed lattice sphere packings}
\author{Yoav Kallus}
\email{ykallus@princeton.edu}
\affiliation{Princeton Center for Theoretical Science, Princeton University, Princeton, New Jersey 08544}
\author{\'Etienne Marcotte}
\affiliation{Department of Physics, Princeton University, Princeton, New Jersey 08544}
\author{Salvatore Torquato}
\affiliation{Department of Physics, Princeton University, Princeton, New Jersey 08544}
\affiliation{Department of Chemistry, Princeton University, Princeton, New Jersey 08544}
\affiliation{Program in Applied and Computational Mathematics, Princeton University, Princeton, New Jersey 08544}
\affiliation{Princeton Institute of the Science and Technology of Materials, Princeton University, Princeton, New Jersey 08544}

\date{\today}

\begin{abstract}
We generate and study an ensemble of isostatic jammed hard-sphere lattices.
These lattices are obtained by compression of a periodic system with an adaptive
unit cell containing a single sphere until the point of mechanical
stability. We present detailed numerical data
about the densities, pair correlations, force distributions, and structure factors
of such lattices. We show that this model retains many of the crucial
structural features of the classical hard-sphere model
and propose it as a model for the jamming and glass transitions
that enables exploration of much higher dimensions than are usually accessible.
\end{abstract}

\maketitle

\section{Introduction}\label{sec:intro}

There are many reasons for studying jammed sphere packings in Euclidean spaces of high dimension. For one,
sphere packing in high dimensions has direct applications for constructing codes in communication and
information theory \cite{SPLAG}. A classical result of Minkowski provides a nonconstructive lower bound
on the optimal packing density in $d$-dimensional Euclidean space, $\varphi > 2^{-d+1}$,
and this lower bound is achieved by a Bravais lattice packing \cite{minkowski}. However,
no general construction is known that achieves this bound in arbitrary dimension, either
by a lattice or a nonlattice packing \cite{SPLAG}. The current asymptotically-best bound is given
by Ball, $\varphi > 2d 2^{-d}$ \cite{ball}. Random processes have been described that seem to satisfy the asymptotic
behavior described by Minkowski's bound, and in some cases to give polynomial corrections to it, 
$\varphi \sim d^\nu 2^{-d}$. Such processes include random sequential addition (RSA) \cite{rsa},
ghost RSA \cite{disorderwins}, and compression of a hard-sphere fluid \cite{skoge,charbikeda}.
These process all result in nonlattice
packings of immense complexity. Here we describe a random process that yields Bravais
lattice packings and that seems to give a larger polynomial correction to Minkowski's bound
than previously described processes, $\varphi \sim d^\nu 2^{-d}$ with $\nu\approx 3$.

Perhaps more importantly, investigating jammed sphere packings in high dimensions has the
clear potential of advancing our fundamental understanding
of granular matter and structural glasses in three dimensions: by going to higher dimensions, not only do we circumvent some of the
technical difficulties encountered in low dimensions -- e.g., the presence of rattlers \cite{skoge} and the
tendency to form crystallites \cite{vanmeel09} -- but we also stand to resolve many ongoing theoretical disputes
if we can determine how crucial quantities like the jamming density scale as a function of dimension
\cite{charbikeda,schmid,ikeda,charbpnas,kurchan}.
However, the increased difficulty of simulating systems in higher dimensions has been a major
roadblock in taking full advantage of their potential to guide the field forward. Pioneering
work has studied jammed sphere packings in up to 6 dimensions \cite{skoge,tj}, and recent heroic efforts
have obtained data for densities in up to 13 dimensions \cite{charbcorwin}. Still, detailed
structural data is only available in up to 10 dimensions \cite{charbcorwin}.
We find that Bravais lattices are simple enough to study in these, and much higher,
dimensions, and yet complex enough that in high enough dimensions they exhibit
all the phenomena of interest in the study of disordered packing.

Jammed configurations of (frictionless) hard spheres are those in which any motion
of the spheres necessary requires an increase in the volume of the system.
These configurations are characterized by mechanical stability: as long
as the volume of the system is held fixed, the system can resist any
set of forces applied to its constituents.
Here we propose to study the jamming behavior of a related system, the hard-sphere lattice model.
A $d$-dimensional lattice sphere packing is a periodic arrangement of non-overlapping spheres in 
Euclidean space $\mathbb{R}^d$ with a single sphere per unit cell, that is, a Bravais lattice (we
will henceforth use the word \textit{lattice} to refer exclusively to Bravais lattices).
In the hard-sphere lattice model, the lattice in question is the dynamic
variable, and the sphere centers occupy all lattice sites. One should not confuse
this kind of lattice model with the more common meaning of \textit{lattice model}, 
where the underlying lattice is fixed and particles move from site to site.
We show that the hard-sphere lattice model allows us to study many of the same phenomena that are
of interest in the classical hard-sphere model, which includes the equilibrium phase diagram, as well as
jammed states, disordered or not. We will henceforth use the modifiers \textit{lattice} and
\textit{classical} to differentiate between phenomena as they occur in the
lattice model and in the classical hard-sphere model. 
With the lattice hard-sphere model, we can study these phenomena in much higher spatial
dimensions than accessible in the classical model:
we obtain and analyze detailed structural data in up to 24 dimensions.
Especially in high dimensions and in the limit $d\to\infty$ there are reasons to expect
some similarities between the lattice and classical hard-sphere models \cite{parisi,decorr}.

The classical hard-sphere model has been invaluable in the study of materials: it is a useful model
for the liquid state \cite{hs-liquid}, the fluid-solid transition \cite{hs-fluidsolid},
the glass transition \cite{hs-glass}, granular flow \cite{hs-granular}, exotic melting behavior
in two dimensions \cite{melt2d}, and countless other phenomena.
Much like the classical hard-sphere model has been incredibly useful in the study of many phenomena,
the recent wave of work on hard-sphere lattices
demonstrates that this versatility by and large carries over to the lattice setting as well: Parisi
established a liquid state theory of lattice sphere packings \cite{parisi}, we have previously
focused on the ground state \cite{marcotte,statlatt}, and here we focus on jamming.

Our results indicate that the typical density of jammed hard-sphere lattices is
likely to be much higher than the density of a classical jammed hard-sphere configuration
in the same dimension. The mean packing fraction of a jammed hard-sphere
lattice appears, based on our analysis (see \ref{sec:res}\ref{ssec:dens}),
to scale asymptotically as $\varphi\sim d^\nu 2^{-d}$, with $\nu=3.01\pm0.01$.
In contrast, estimates and predictions for classical jammed hard spheres give asymptotic
scaling with an exponent $1\le\nu\le2$ \cite{jin,charbikeda}.
The fact that jammed hard-sphere lattices seem to be much denser than
classical jammed configurations might come as a surprise, because in the case of the optimal packing,
the densest configurations are expected to be much more dense than
the densest (Bravais) lattice configurations in most high-enough dimensions \cite{disorderwins}.
However, the fact that jammed lattices must have at least $d(d+1)$ contacts
around each sphere, compared to an average of $2d$ contacts in a classical
jammed configuration (see \ref{sec:thy}), suggests that we might also expect higher densities.

We report here pair correlation
functions and contact force distributions similar to the ones observed in the
classical setting. In particular, we observe power law behaviors for the distribution
of weak contact forces and for the near-contact pair correlation. The power
law exponents do not depend on dimension, but they are
different than those observed in the classical model \cite{charbcorwin}.
In \ref{sec:thy}, we will give some background and formulate the lattice sphere packing problem as an optimization problem.
This formulation gives rise to the description of contact forces in a jammed packing as Lagrange
multipliers. In \ref{sec:methods}, we will describe the procedure by which we construct ensembles
of isostatic extreme lattices in dimensions $15$ to $24$. In \ref{sec:res}, we describe
our observations of that ensemble, summarized above. We end with some closing remarks in
\ref{sec:cnc}.

\section{Theory}\label{sec:thy}

The study of lattices as a special case of sphere packing in arbitrary dimensions has
a long and celebrated history \cite{SPLAG,thompson}. In the physical setting,
lattices have been considered as ordered phases of hard-sphere systems, but the
behavior of a
system restricted to take only lattice configurations has only recently begun
to draw serious study \cite{parisi}.
Parisi considered the problem of calculating the entropy of such
a system as a function of unit cell volume. He uses results of Rogers as a starting
point to derive the basic tools and techniques needed to establish a formal statistical
mechanics theory of lattice sphere packings \cite{parisi}.
The second- and third-named authors studied
the behavior of the system under rapid compression using the Torquato-Jiao
algorithm \cite{tj,marcotte}.
The first-named author performed quasistatic compression simulations using a Monte Carlo method,
which showed that the system experiences a fluid-solid
crystallization transition in moderately high dimensions \cite{statlatt}.

The use of the terms \textit{fluid} and \textit{crystallization}
might be confusing or seem inappropriate when all configurations considered are
\textit{a priori} lattices and therefore ostensibly crystalline. However,
in light of the fact that Parisi establishes a virial expansion for lattice
sphere packings, there is much sense in describing a fluid phase as the range
of validity of this expansion. Moreover, as Parisi shows, for some purposes as $d\to\infty$
the distribution of lattice points is well-approximated by an uncorrelated distribution
of random points (see Section 4.4 of Ref.\@ \cite{parisi}). The so-called solid phase
is associated with the densest lattice packing in a given dimension, the ground state
of the system.

It is important to note here that while we can describe each configuration as
a periodic pattern occupying the entire $d$-dimensional Euclidean space, this pattern
is always composed of many copies of a single unit cell. Therefore, the system
should be considered a spatially nonextended system in the sense that all its degrees
of freedom are restricted to a single unit cell and once this unit cell is defined
the entire periodic pattern in $\mathbb{R}^d$ is prescribed. For any fixed value of $d$,
the number of parameters required to fully describe the configuration of the system is finite,
despite the fact that the positions of an infinite number of spheres are thus described.
Therefore, at any fixed $d$, the system does not have a thermodynamic limit, and any phase
transition is in fact rounded. The role of the thermodynamic limit is filled by $d\to\infty$.
This situation bares a resemblance to mean-field spin glass models where every pair (or p-tuple, as the case may be)
of spins interact, such as the Sherrington-Kirkpatrick spin glass or the spherical p-spin glass
\cite{pedestrian}. 

As with classical hard spheres, the fluid phase remains metastable at densities above
the crystallization point, and if the system is compressed quickly enough, it will reach
a mechanically stable, jammed, structure without falling into the solid phase. In fact, in high
dimensions, even very slow compression will lead to this result. According to the Random-First Order Transition
(RFOT) theory, the system does experience a fluid-solid transition, but this solid
phase is associated with one of many local maxima of the density, and not the global maximum \cite{parisi10}.
At a finite pressure, the system will occupy many configurations in the basin of attraction
around the local maximum, but as the pressure approaches infinity, the system configuration
will approach the local maximum itself.
Lattices that are local maxima of density in the space of valid packings are known in the literature
as extreme lattices, and it is known that the number of contacts around
any sphere in an extreme lattice cannot be smaller than $z=d(d+1)$ \cite{perfect}.
This fact is similar, and indeed has similar origins, to the \textit{isostatic} condition in the classical setting,
which requires at least $z=2d$ contacts on average around each sphere. 
Classical jammed sphere packings are observed to also have no more than this number of contacts \cite{doneviso}.
In Ref.\@ \cite{marcotte}, the authors identify tens of thousands of extreme lattices in dimensions $d\le19$ as
the final states of their algorithm. They find that an overwhelming majority in dimensions $d\ge 17$ have
exactly $d(d+1)$ contacts. We call such lattices isostatic, and this notion of isostaticity should be
contrasted with the classical notion of isostaticity, where an average of $2d$ contacts are incident on each sphere.

The lattice sphere packing problem can be formulated as the following optimization problem over
symmetric $d\times d$ matrices:
\begin{equation}
    \begin{aligned}
	\text{minimize} & \hspace{0.5em} \det G\text, \\
	\text{subject to} 
	& \hspace{0.5em} 
	\langle \mathbf{n}, G\mathbf{n} \rangle \ge 1 \text{ for all }\mathbf{n}\in\mathbb{Z}^d\setminus\{0\}\text.
    \end{aligned}
    \label{eqn:optim}
\end{equation}
Note that any feasible matrix $G$ must be positive definite (depending on context, we will use the equivalent terms \textit{feasible}
which comes from the optimization literature and \textit{admissible} from the geometry literature).
A feasible matrix $G$ corresponds to a packing of unit-diameter spheres, centered at the lattice points
$M\mathbf{n}$, $\mathbf{n}\in\mathbb{Z}^d$, where $M$ is some matrix such that $G=M^T M$
($M$ is determined by $G$ only up to rotation). The choice of unit diameters
fixes our scale for distances and wavenumbers, which will take dimensionless values
in units of the sphere diameter and its inverse respectively.
We call $G$ the \textit{Gram matrix} of the lattice,
and $M$ the \textit{generating matrix}. The fraction of space filled by the packing is $\varphi=2^{-d} (\det G)^{-1/2} V_d$,
where $V_d=\pi^{d/2}/\Gamma(\tfrac{d}{2}+1)$ is the volume of a unit-\textit{radius} ball in $d$ dimensions.
Note that the constraints in \ref{eqn:optim} actually come in equivalent pairs, since
$\langle \mathbf{n}, G\mathbf{n}\rangle =\langle -\mathbf{n}, G(-\mathbf{n}) \rangle$.

Each \textit{active} constraint (a constraint
satisfied with equality), labeled by $\mathbf{n}$, corresponds to one contact
per unit cell of the packing: namely between the sphere
centered at $M \mathbf{n'}$ and that at $M (\mathbf{n'}+\mathbf{n})$ for any $\mathbf{n'}\in\mathbb{Z}^d$.
We associate a contact force to this contact using the Lagrangian formulation of \ref{eqn:optim}.
A stationary point of the Lagrangian satisfies
\begin{equation}
    (\det G) G^{-1} = \sum_{i=1}^z f_i \mathbf{n}_i \mathbf{n}_i^T\text,
    \label{eqn:fb}
\end{equation}
where the sum runs over all contacts $\mathbf{n}_i$, $i=1,\ldots,z$, and we assume that $f_i=f_{i'}$ when $\mathbf{n}_i=-\mathbf{n}_{i'}$ due
to Newton's third law. Note that in contrast with the classical setting, this is not the same as simple mechanical force balance,
since here we have force balance automatically from the equality of forces along contacts at antipodal points on the surface
of the sphere. For stability, the forces must be repulsive $f_i>0$.
The criterion \ref{eqn:fb} with $f_i>0$ is known in the literature as \textit{eutaxy} \cite{perfect}.
It turns out that an additional criterion
for stability is that the rank-1 matrices $\mathbf{n}_i \mathbf{n}_i^T$ have to form a complete basis of the
space of symmetric $d\times d$ matrices, a criterion known as \textit{perfection} \cite{perfect}.
An equivalent definition of perfection is that the Gram matrix is fully determined by the
identity of the contacts $\mathbf{n}_i$, that is, there is a unique Gram matrix such that 
$\langle \mathbf{n}_i,G\mathbf{n}_i\rangle=1$ for all $i=1,\ldots,z$.
An admissible lattice which satisfies these two mechanical stability criteria is an extreme lattice, a local minimum
of the optimization problem \ref{eqn:optim}. The minimum number of contacts which makes this
possible is the isostatic number $z=d(d+1)$. We limit our attention from
this point on to isostatic extreme lattices.
The average contact force can be obtained from \ref{eqn:fb} by multiplying both sides by $G$ and taking the trace. Using the fact that
$\trc \mathbf{n}_i\mathbf{n}_i^T G = \langle \mathbf{n}_i, G \mathbf{n}_i\rangle = 1$, we have that $\sum f_i = d (\det G)$ and
so $\langle f\rangle = (\det G)/(d+1)$.

We will be interested in the distribution of distances between pairs of spheres in the lattice.
Consider the function 
\begin{equation*}
    Z(r)=|\{\mathbf{n}\in\mathbb{Z}^d : \langle \mathbf{n},G\mathbf{n}\rangle \le r^2\}|-1\text,
\end{equation*}
which gives the number of spheres whose center is a distance
of $r$ or less from the center of some fixed sphere, not counting that sphere.
This function is related to the more common pair correlation function
$g(r) = (dZ/dr) / (d 2^d r^{d-1} \varphi)$.
The problem of finding the number of vectors in a lattice shorter than some length,
given its generating matrix $M$ or its Gram
matrix $G$, is well studied and is known to be NP-hard \cite{nphard}. 

The structure factor of the packing $S(\mathbf{k})$ is given simply by the reciprocal lattice $M^{-1}\mathbb{Z}^d$:
\begin{equation*}
    S(\mathbf{k}) = \det{M^{-1}}
    \sum_{\mathbf{m}\in \mathbb{Z}^d} \delta (\mathbf{k}-M^{-1} \mathbf{m})\text.
\end{equation*}
Each nonzero point $\mathbf{k}$ of the reciprocal lattice, if it is ``visible'' from the origin, corresponds
to a partition of the direct lattice points into planes separated by a distance $1/|\mathbf{k}|$.
Note that extremity imposes a constraint on the length of the shortest reciprocal lattice vector:
a sphere in a given plane $\langle \mathbf{k},\mathbf{x}\rangle=0$ must be in contact with
at least $d$ spheres in the half-space $\langle \mathbf{k},\mathbf{x}\rangle<0$, or else
the planes defined by $\mathbf{k}$ can be brought closer together and the lattice is not locally
optimal in density. The largest possible separation along the direction of $\mathbf{k}$ between the 
original sphere and the closest of the touching spheres is obtained when the spheres
form a regular simplex, and the separation between the layers is $\sqrt{(1+\tfrac{1}{d})/2}$.
Therefore $|\mathbf{k}|\ge\sqrt{2d/(d+1)}$ for any nonzero reciprocal
lattice vector $\mathbf{k}$. This bound is achieved uniquely by the shortest reciprocal lattice vectors
for the direct lattice generated by the root system $A_d$, which is in fact extreme
and isostatic \cite{SPLAG}.

\section{Methods}\label{sec:methods}

In Ref.\@ \cite{marcotte} 10,000 lattice packings were generated in each dimension
$d=10,\ldots,18$ and 100,000 in $d=19,20$. Using the same procedure, we generate
10,000 additional lattices in each dimension $d=21,\ldots,24$, and in dimensions
$d=19,20$ we use only 20,000 of the 100,000 generated in Ref.\@ \cite{marcotte}.
The method used seeks to optimize the density using a sequence of moves, leading
in some cases to the globally optimal solution and in other cases to a local optimum.
Therefore, we expect the lattices obtained to be locally
optimal, that is, extreme, but perhaps affected by numerical errors. For each lattice we calculate
$Z(1+5\times10^{-8})$, the number of approximate contacts, and keep a record
of the corresponding integer vectors $\mathbf{n}$. We keep all the lattices
for which this number is exactly $d(d+1)$. Next, we reconstruct the Gram matrix
from the identity of the contacts, which is guaranteed to be possible by perfection.
However, due to numerical errors, for a small number of lattices the rank-1 matrices do not form
a complete basis, and the reconstruction fails. We reject these lattices.
Note that since we are using the integer vectors to determine the Gram matrix,
we can achieve arbitrary precision, but for the purposes of this article, the precision
of double precision floating point numbers suffices. We recalculate the number of neighbors
using the precise Gram matrix, rejecting the small number of lattices that are not, after all,
isostatic. Finally we calculate the contact
forces by solving \ref{eqn:fb}. Again, due to numerical errors in the original
data we have to reject a small number of lattices that yield negative forces.
The number of extreme isostatic lattices we end up with is given in \ref{tab:num-iel}.
For $d\le14$, only a small fraction of lattices generated are isostatic, but this fraction
quickly grows to dominate the ensemble in higher dimensions. We suspect that
this fraction should theoretically tend to one, but does not in our case because of
numerical errors that make it difficult to exactly identify contacts.
In \ref{sec:res} we study the isostatic extreme lattices only in the dimensions
where they constitute the majority of lattices generated, namely $d\ge15$.

Our method of obtaining an ensemble of extreme lattices should be compared,
perhaps, to the method used in Ref.\@ \cite{extremesamp}. There, the authors
generate a large number of extreme lattices (tens of thousands) in dimensions
$d\le13$. However, in higher dimensions, their method becomes inefficient and
a much smaller number of distinct extreme lattices is generated (a few hundreds
to a few thousands) in dimensions $14\le d\le19$. A major issue in Ref.
\cite{extremesamp} is that the method samples the same lattice multiple
times, and therefore a large sample may contain only a small number of
distinct lattices. With our methods we have no such issue: we find that
any two lattices among our sample of extreme isostatic lattices in each
dimension $15\le d\le24$ have distinct values of the determinant, which
implies necessarily that they are distinct. The two methods appear to
complement each other well: the method of Ref.\@ \cite{extremesamp} is
well-suited to generate large samples of distinct extreme lattices
in dimensions $d\le13$, where the present method would produce many
repetitions of the densest lattices, while the present method is
well-suited for dimensions $d\ge14$, where it produces hardly any
repetitions among the isostatic extreme lattices.

\begin{table}
    \centering
	\begin{filecontents*}{counts.dat}
13	10000	    365
14	10000	   1625
15	10000	   5196 
16	10000	   6761 
17	10000	   9235 
18	10000	   9590 
19	20000	  19200 
20	20000	  19085 
21	10000	   9473 
22	10000	   9406 
23	10000	   9281 
24	10000	   9205 
\end{filecontents*}
\pgfplotstabletypeset[
	    columns/0/.style={ column name = {$d$}, column type = {c} },
	    columns/1/.style={ column name = {Runs}, column type = {c} },
	    columns/2/.style={ column name = {Extreme Isostatic}, column type = {c} },
	    every head row/.style={before row={\hline \hline}, after row=\hline},
	    every last row/.style={after row={\hline \hline}}
	]{counts.dat}
	\caption{Number of runs of the algorithm of \cite{marcotte} used in each
	dimension $d$, and the number
	of extreme isostatic lattice obtained as a result of processing the resulting lattices
	according to the procedure outlined in \ref{sec:methods}.}
    \label{tab:num-iel}
\end{table}

\section{Results}\label{sec:res}

\subsection{Densities} \label{ssec:dens}

\begin{figure}
    \centering
    \scalebox{0.8}{
	\begin{tikzpicture}
	    \begin{axis}[
		xlabel={$d$}, ylabel={$2^d\langle\varphi\rangle$},
		domain=15:24, xmin=14.5, xmax=24.5,
		ymin=0, ymax=900,
		y label style={at={(0.05,0.5)}},
	    ]
		\addplot[red, mark=+, only marks] table[x index=0,y index=1] {phid.dat};
		\addplot[] {0.0567159 * x^3.01226};
	    \end{axis}
	    \begin{loglogaxis}[
		width=\myinsetwidth,domain=15:24,
		xshift=0.05\textwidth, yshift=3.5cm, xmin=14.5, xmax=25.5,
		ymin=180, ymax=1000,
		xticklabels = {$15$,$20$,$25$}, yticklabels = {$200$,$500$,$1000$},
		xtick = {15,20,25}, ytick = {200,500,1000}
	    ]
		\addplot[red, mark=+, only marks] table[x index=0,y index=1] {phid.dat};
		\addplot[] {0.0567159 * x^3.01226};
	    \end{loglogaxis}
	\end{tikzpicture}
    }
	\caption{
	    The mean density $\langle\varphi\rangle$ of jammed isostatic lattice sphere packings as a
	    function of dimension, rescaled by $2^d$. The line is the best fit power law
	    $2^d\langle\varphi\rangle = 0.0567 d^{3.01}$ and the inset shows the same
	    on a log-log scale.
	}
    \label{fig:phi-iel}
\end{figure}
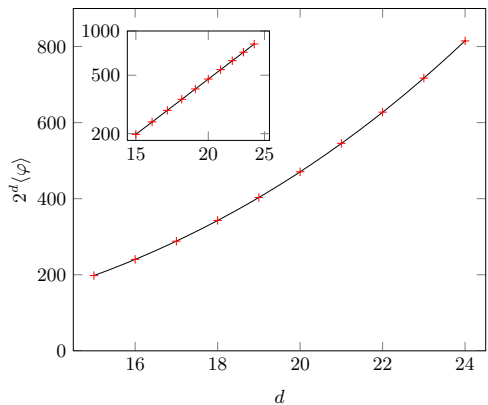

Understanding the distribution of densities at which hard spheres jam and its
dependence on dimension is one of the major challenges facing the field.
While it is now widely accepted that hard spheres can jam at a range of 
different densities \cite{parisi10,torqstil10,chaudhuri10,nonuniversality},
different causes have been proposed to explain this
variability: for example, variability in the thermodynamic parameters
of the initial configurations \cite{liu-review}, or the existence of crystalline
regions, detectable using a carefully constructed order parameter \cite{mrj}.
For a finite system, we always expect to observe a distribution of
jamming densities. As the size of the system grows and if we use
a fixed algorithm or experimental protocol to generate the packing, the distribution
is expected to narrow and tend to a delta function as the thermodynamic
limit is approached \cite{ohern03}. In general, the limit density depends
on the algorithm used.

Additionally, different theories predict different scalings for the typical
jamming density as a function of dimension. A local geometric argument based on
a simple assumption for the pair correlation predicts $\varphi\sim d 2^{-d}$
\cite{jin}. A replica theory (RT) calculation also predicts a
density $\varphi_\text{th}\sim d 2^{-d}$ \cite{parisi10}, while a mode-coupling-theory (MCT)
calculation predicts a MCT transition at a density $\varphi_\text{MCT} \sim d^2 2^{-d}$,
implying that jamming must occur at an even higher density \cite{schmid,ikeda,charbikeda}.
While our present results do not address any of these controversies directly,
being obtained for a different system, we hope that they will nevertheless
be useful in resolving them by providing both intuition and a convenient testing
ground for theories.

\begin{figure}
    \centering
    \scalebox{0.8}{
	\begin{tikzpicture}
	    \begin{axis}[
		xlabel={$\varphi/\langle\varphi\rangle$},
		ylabel={$\langle\varphi\rangle P(\varphi)$},
		domain=0.85:1.15, xmin=0.85, xmax=1.15,
		y label style={at={(0.05,0.5)}},
		legend style={at={(0.03,0.97)},anchor=north west},
		legend entries={$15$,$16$,$17$,$18$,$19$,$20$,$21$,$22$,$23$,$24$},
		extra y ticks={0.},
		extra y tick style={grid=major},
	    ]
		\pgfplotsset{cycle list name=list1};
		\addplot table[x index=0,y index=1] {dens.dat};
		\addplot table[x index=0,y index=2] {dens.dat};
		\addplot table[x index=0,y index=3] {dens.dat};
		\addplot table[x index=0,y index=4] {dens.dat};
		\addplot table[x index=0,y index=5] {dens.dat};
		\addplot table[x index=0,y index=6] {dens.dat};
		\addplot table[x index=0,y index=7] {dens.dat};
		\addplot table[x index=0,y index=8] {dens.dat};
		\addplot table[x index=0,y index=9] {dens.dat};
		\addplot table[x index=0,y index=10] {dens.dat};
		\addplot[thick] {exp(3.2566573758632695 - 534.121296209138*(x-1)^2)};
	    \end{axis}
	\end{tikzpicture}
    }
	\caption{
	    The distribution of densities, rescaled by the mean density. The line is
	    the best overall Gaussian fit.
	}
    \label{fig:dens}
\end{figure}
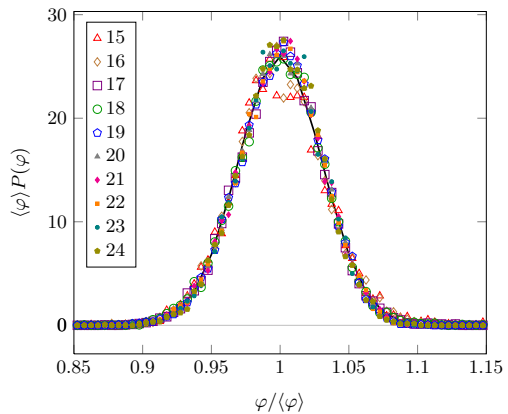

The mean lattice density $\langle\varphi\rangle$ as function of dimension is plotted in \ref{fig:phi-iel}.
The dependence on the dimension is well-fit by $\langle\varphi\rangle \simeq c d^{\nu} 2^{-d}$,
with $\nu=3.01\pm0.01$. These densities are much higher than the jamming densities
for classical hard spheres, and the relative difference is expected to widen with dimension.
This result seems to be in stark contrast to the situation expected for
the densest, rather than jammed, packings: in high enough dimensions, the densest 
packing of equal sized hard spheres is expected to be much denser
than the densest (Bravais) lattice packing \cite{disorderwins}.
If the RT and MCT calculations can be carried over to the
lattice setting, it would be interesting to see if any of them are contradicted
by these results. The geometric calculation of Ref.\@ \cite{jin}
predicts a scaling $\varphi\sim z 2^{-d}$, which in this case
would imply $\varphi\sim d^2 2^{-d}$.

In \ref{fig:dens} we plot the distribution of densities in each dimension rescaled
by the mean density in that dimension. The distributions appear nearly identical,
seemingly contradicting the expectation that the distribution should narrow as the system
size grows. However, we note that the standard deviation of the distribution does show a slow
narrowing trend: from $3.3\times 10^{-2}$ for $d=15$ to $2.9\times 10^{-2}$ for $d=24$.
We cannot predict from the current data whether the distribution
will tend toward a delta function in the limit $d\to\infty$.
We also note that the algorithm used here for different dimensions could arguably 
be considered not fixed, since it is not clear what the correct way to scale with
dimension such parameters as the step size and the influence radius (see Ref.
\cite{marcotte}). However, the persistence of a finite range of relative jamming densities
in the thermodynamic limit even for a fixed algorithm might be possible due to the
fact that the model has no spatial extent and is therefore not required to be
self-averaging \cite{pedestrian}.

\subsection{Pair correlation}

\begin{figure}
    \centering
    \scalebox{1.0}{
	\begin{tikzpicture}
	    \begin{axis}[
		xlabel={$r$},
		ylabel={$g(r)$},
		domain=1.:1.8, xmin=1., xmax=1.8,
		ymin=0.85,ymax=1.15,
		y label style={at={(0.05,0.5)}},
		legend style={at={(1.00,0.00)},anchor=south east},
		legend entries={$15$,$16$,$17$,$18$,$19$,$20$,$21$,$22$,$23$,$24$},
		legend columns=3, 
		legend style={ /tikz/column 2/.style={column sep=5pt}, },
		legend style={ /tikz/column 3/.style={column sep=5pt}, },
		extra x ticks={1.41421, 1.73205},
		extra x tick style={grid=major},
		extra x tick labels={},
		extra y ticks={1.},
		extra y tick style={grid=major},
	    ]
		\pgfplotsset{cycle list name=list2};
		\addplot table[x index=0,y index=1] {qg15.dat};
		\addplot table[x index=0,y index=1] {qg16.dat};
		\addplot table[x index=0,y index=1] {qg17.dat};
		\addplot table[x index=0,y index=1] {qg18.dat};
		\addplot table[x index=0,y index=1] {qg19.dat};
		\addplot table[x index=0,y index=1] {qg20.dat};
		\addplot table[x index=0,y index=1] {qg21.dat};
		\addplot table[x index=0,y index=1] {qg22.dat};
		\addplot table[x index=0,y index=1] {qg23.dat};
		\addplot table[x index=0,y index=1] {qg24.dat};
	    \end{axis}
	\end{tikzpicture}
    }
	\caption{
	    The average pair correlation $g(r)$ in a jammed isostatic
	    lattice of dimension $d$, where $d=15,\ldots,24$.
	    The vertical lines mark the apparent singularities
	    at $r=\sqrt{2}$ and $r=\sqrt{3}$. The distances are
	    given in units of the sphere diameter.
	}
    \label{fig:ztil}
\end{figure}
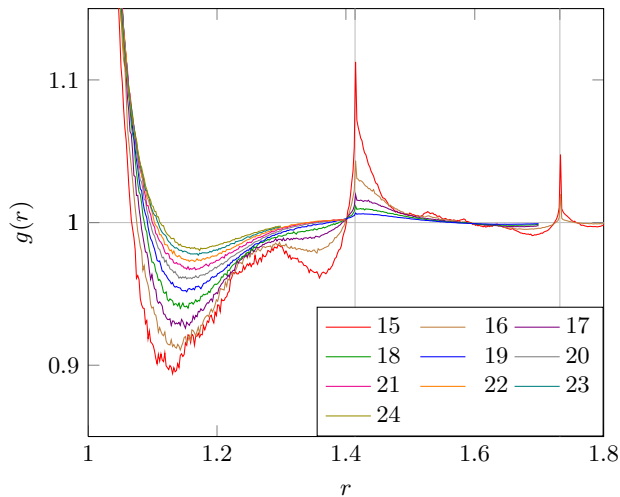

The pair correlation function $g(r)$, averaged over the jammed isostatic lattices
in each dimension, is plotted in \ref{fig:ztil}. The pair correlation is
zero for $r<1$ due to the hard-sphere constraint and has a delta-function
singularity at $r=1$ due to the pairs that are in contact. For $r>1$ we
observe a number of distinctive features: a power-law divergence $g(r)\sim r^{-\gamma}$ for $r\to1^+$,
and apparent singularities at $r=\sqrt{2}$ and $r=\sqrt{3}$. At finite-pressure
the delta-function singularity at $r=1$ is also expected to broaden into a power-law,
whose exponent is related to the distribution of contact forces
and which will control the behavior at $r\to1^+$
\cite{donev05,charbcorwin}. Since we are working with configurations at
infinite pressure, the two singular behaviors at $r=1$ are completely
separated, and can be studied separately in the pair-correlation $g(r)$
and in the contact force distribution (see \ref{sec:res}\ref{ssec:frc}).
Note that we were able to easily approach the infinite pressure
limit in this system because a simple linear equation gives
the exact Gram matrix once the contacts are identified. In the classical
setting, the equations determining the configuration based on the identity
of contacts are nonlinear, and highly nontrivial to solve. This is another
remarkable advantage of the lattice model.

We observe that the features of $g(r)$, especially the singularities at $r=\sqrt{2}$ and $r=\sqrt{3}$,
become less pronounced in higher dimensions, and the correlation
function approaches the asymptotic value $g(\infty)=1$ faster.
The weakening of correlation features with increased dimension
is a prediction of the principle of decorrelation \cite{disorderwins}
This principle has been shown to apply not only to amorphous configurations but
also to lattices \cite{decorr}.

While, for any individual lattice, $g(r)$ is, strictly speaking,
a sum of delta functions, the spacing between these is much smaller than the
sphere radius, and $g(r)$ can be considered a continuous curve
(apart from the aforementioned singularities). Therefore, the
curves in \ref{fig:ztil}, which are averaged over many lattices,
can also be considered to represent the typical lattice.
Note that this is a non-trivial feature of this ensemble of lattices.
The extreme lattices that are densest, or close to densest, in any
dimension are typically given by Gram matrices whose
elements are rational numbers of small denominator \cite{SPLAG,marcotte,statlatt}.
Therefore, the possible squared pair separations
$r_{ij}^2=\langle{\mathbf{n}_i-\mathbf{n}_j,G(\mathbf{n}_i-\mathbf{n}_j)\rangle}$
must also be rational numbers of small denominator, and
the pair correlation function will consist, at least for $r\lesssim 2$,
of only a few delta functions.

\begin{figure}
    \centering
    \scalebox{0.8}{
	\begin{tikzpicture}
	    \begin{axis}[
		xlabel={$d$}, ylabel={$A$},
		domain=15:24, xmin=14.5, xmax=24.5,
		ymin=0, ymax=60,
		y label style={at={(0.05,0.5)}},
	    ]
		\addplot[red, mark=+, only marks] table[x index=0,y index=1] {Ad.dat};
		\addplot[] {0.00143089*x^3.29828};
	    \end{axis}
	    \begin{loglogaxis}[
		width=\myinsetwidth,domain=15:24,
		xshift=0.05\textwidth, yshift=3.5cm, xmin=14.5, xmax=25.5,
		ymin=8, ymax=60,
		xticklabels = {$15$,$20$,$25$}, yticklabels = {$10$,$20$,$50$},
		xtick = {15,20,25}, ytick = {10,20,50}
	    ]
		\addplot[red, mark=+, only marks] table[x index=0,y index=1] {Ad.dat};
		\addplot[] {0.00143089*x^3.29828};
	    \end{loglogaxis}
	\end{tikzpicture}
    }
	\caption{
	    The amplitude of the near-contact singularity in the pair correlation
	    as a function of dimensions. The line is the best fit power law
	    $A = (1.43\times 10^{-3})d^{3.30}$. The inset shows the same on
	    a log-log scale.
	}
    \label{fig:amplitude}
\end{figure}
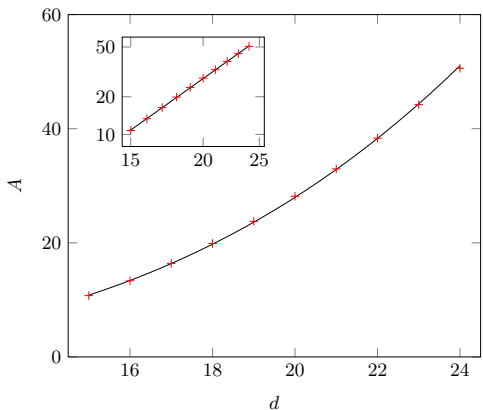

The divergence for $r\to1^+$
remains a strong feature for all the dimensions studied here.
Such a divergence, described by a power law, is also observed
in classical jammed packings \cite{donev05,silbert06}, and the power law
exponent seems to be independent of dimension \cite{charbcorwin}.
We use logarithmic binning
to extract the correlation of near-contacting pairs, which appears
to be well-described by a power law
$\langle Z(1+\xi)\rangle \simeq A \xi^{1-\gamma}$ (see \ref{fig:z-plaw}).
The best-fit values of $\gamma$ are given in \ref{tab:exps}.
The value of the power law exponent seems to take a dimension-independent
value of $\gamma=0.314\pm0.004$.
The amplitude of the near-contact singularity, on the other hand, seems
neither to be a constant with dimension nor to be directly proportional
to the density.
Instead, the amplitude seems to scale with a distinct power law exponent
as a function of dimension, $A\simeq c d^{\tilde{\nu}}$, where
$\tilde{\nu}=3.30\pm0.05$ (see \ref{fig:amplitude}).

\begin{table}
    \centering
	\begin{filecontents*}{exps2.dat}
15	0.31275236479702373	0.3534470660144051
16	0.31365619847721826	0.36854841550755313
17	0.3202509707197303	0.3633266613438857
18	0.31766625838328755	0.3603198303772226
19	0.31718339750543223	0.36535897538078355
20	0.3159231566612183	0.37311026774334666
21	0.313449712055047	0.38340405045488213
22	0.3083142330090469	0.3856060849850702
23	0.3096986278465603	0.3829832681365626
24	0.3070857906388309	0.3775234073204141
\end{filecontents*}
\pgfplotstabletypeset[
	    columns/0/.style={ column name = {$d$}, column type = {c} },
	    columns/1/.style={ column name = {$\gamma$}, precision=4, zerofill, column type = {c} },
	    columns/2/.style={ column name = {$\theta$}, precision=4, zerofill, column type = {c} },
	    every head row/.style={before row={\hline \hline}, after row=\hline},
	    every last row/.style={after row={\hline \hline}}
	]{exps2.dat}
	\caption{The value in each dimension of the best-fit power law exponents for the
	    data shown in \ref{fig:z-plaw} and \ref{fig:force-plaw}.}
    \label{tab:exps}
\end{table}

\begin{figure}
    \centering
    \scalebox{0.8}{
	\begin{tikzpicture}
	    \begin{loglogaxis}[
		xlabel={$r-1$},
		ylabel={$(r-1)(dZ/dr)$},
		domain=0.0001:0.005, xmin=0.0001, xmax=0.005, ymax=1.5,
		y label style={at={(0.02,0.5)}},
		legend style={at={(1.00,0.00)},anchor=south east},
		legend entries={$15$,$16$,$17$,$18$,$19$,$20$,$21$,$22$,$23$,$24$},
		legend columns=3, 
		legend style={ /tikz/column 2/.style={column sep=5pt}, },
		legend style={ /tikz/column 3/.style={column sep=5pt}, },
	    ]
		\pgfplotsset{cycle list name=list2};
		\addplot table[x index=0,y index=1] {z-loglog.dat};
		\addplot table[x index=0,y index=2] {z-loglog.dat};
		\addplot table[x index=0,y index=3] {z-loglog.dat};
		\addplot table[x index=0,y index=4] {z-loglog.dat};
		\addplot table[x index=0,y index=5] {z-loglog.dat};
		\addplot table[x index=0,y index=6] {z-loglog.dat};
		\addplot table[x index=0,y index=7] {z-loglog.dat};
		\addplot table[x index=0,y index=8] {z-loglog.dat};
		\addplot table[x index=0,y index=9] {z-loglog.dat};
		\addplot table[x index=0,y index=10] {z-loglog.dat};
		\addplot[densely dashed] {55. * x^0.686402};
		\addplot[densely dashed] {9.5 * x^0.686402};
	    \end{loglogaxis}
	\end{tikzpicture}
    }
	\caption{
	    Logarithmically binned histogram of pair separations.
	    The curves are fit by power laws $c (r-1)^{1-\gamma}$ with exponents
	    given in \ref{tab:exps}. The dashed lines have a slope of 
	    $0.686$, corresponding to the average value of $1-\gamma$.
	}
    \label{fig:z-plaw}
\end{figure}
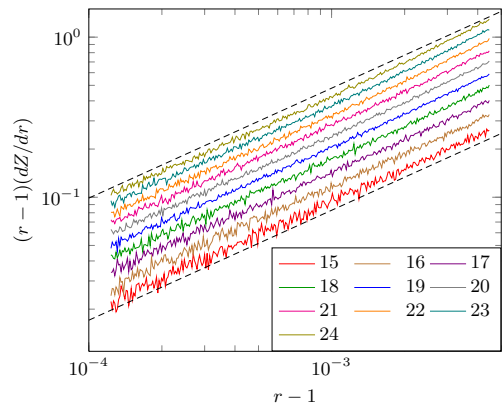

\subsection{Forces}\label{ssec:frc}

\begin{figure}
    \centering
    \scalebox{0.8}{
	\begin{tikzpicture}
	    \begin{axis}[
		xlabel={$f/\langle f\rangle$},
		ylabel={$P(f/\langle f\rangle)$},
		domain=0.:3.5, xmin=0., xmax=3.5, ymin=0.,
		y label style={at={(0.05,0.5)}},
		legend style={at={(0.09,0.03)},anchor=south west},
		legend entries={$15$,$16$,$17$,$18$,$19$,$20$,$21$,$22$,$23$,$24$},
	    ]
		\pgfplotsset{cycle list name=list2};
		\addplot table[x index=0,y index=1] {force.dat};
		\addplot table[x index=0,y index=2] {force.dat};
		\addplot table[x index=0,y index=3] {force.dat};
		\addplot table[x index=0,y index=4] {force.dat};
		\addplot table[x index=0,y index=5] {force.dat};
		\addplot table[x index=0,y index=6] {force.dat};
		\addplot table[x index=0,y index=7] {force.dat};
		\addplot table[x index=0,y index=8] {force.dat};
		\addplot table[x index=0,y index=9] {force.dat};
		\addplot table[x index=0,y index=10] {force.dat};
	    \end{axis}
	    \begin{semilogyaxis}[
		width=\myinsetwidth,
		xshift=0.225\textwidth, yshift=3.5cm,
		domain=1.2375:4., xmin=2.5, xmax=4.,
		ymin=0.0001
	    ]
		\pgfplotsset{cycle list name=list2};
		\addplot table[x index=0,y index=1] {force-tail.dat};
		\addplot table[x index=0,y index=2] {force-tail.dat};
		\addplot table[x index=0,y index=3] {force-tail.dat};
		\addplot table[x index=0,y index=4] {force-tail.dat};
		\addplot table[x index=0,y index=5] {force-tail.dat};
		\addplot table[x index=0,y index=6] {force-tail.dat};
		\addplot table[x index=0,y index=7] {force-tail.dat};
		\addplot table[x index=0,y index=8] {force-tail.dat};
		\addplot table[x index=0,y index=9] {force-tail.dat};
		\addplot table[x index=0,y index=10] {force-tail.dat};
	    \end{semilogyaxis}
	\end{tikzpicture}
    }
	\caption{
	    The distribution of contact forces. The behavior of the tail
	    of the distribution is shown in the inset on a semi-log plot.
	}
    \label{fig:forc}
\end{figure}
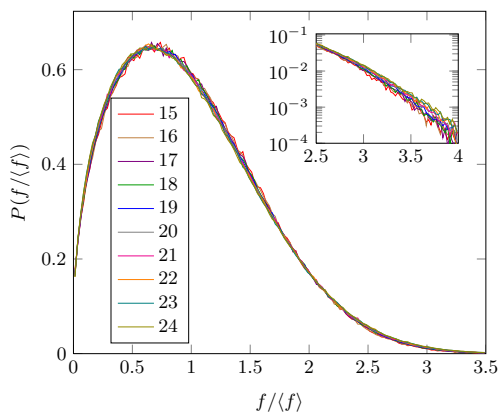

The distribution of the contact force at a random contact in a lattice,
rescaled by the mean value for that lattice, is plotted in \ref{fig:forc}.
The mean value for each lattice is determined by its
density (see \ref{sec:thy}). Again we note that for any individual
lattice, this distribution is discrete, but that it approaches
a smooth distribution as the number of contacts increases with
dimension. For the densest lattices in any dimension, this is again not
the case: not only are the contact forces not well-defined (because of
hyperstaticity), but the high degree of symmetry implies that
there are only a few inequivalent classes of contacts \cite{SPLAG,extremesamp}.

\begin{figure}
    \centering
    \scalebox{0.8}{
	\begin{tikzpicture}
	    \begin{loglogaxis}[
		xlabel={$f/\langle f\rangle$},
		ylabel={$(f/\langle f\rangle) P(f/\langle f\rangle)$},
		domain=0.00123:0.35, xmin=0.001, xmax=0.45,
		y label style={at={(0.02,0.5)}}
	    ]
		\pgfplotsset{cycle list name=list2};
		\addplot table[x index=0,y index=1] {force-loglog2.dat};
		\addplot table[x index=0,y index=2] {force-loglog2.dat};
		\addplot table[x index=0,y index=3] {force-loglog2.dat};
		\addplot table[x index=0,y index=4] {force-loglog2.dat};
		\addplot table[x index=0,y index=5] {force-loglog2.dat};
		\addplot table[x index=0,y index=6] {force-loglog2.dat};
		\addplot table[x index=0,y index=7] {force-loglog2.dat};
		\addplot table[x index=0,y index=8] {force-loglog2.dat};
		\addplot table[x index=0,y index=9] {force-loglog2.dat};
		\addplot table[x index=0,y index=10] {force-loglog2.dat};
		\addplot[thick] {0.4296293018215329 * x^1.3713628027264124};
	    \end{loglogaxis}
	\end{tikzpicture}
    }
	\caption{
	    Logarithmically binned histogram of contact forces.
	    The black line shows the best fit power law
	    $(f/\langle f\rangle) P(f/\langle f\rangle) = 0.430 (f/\langle f\rangle)^{1.371}$.
	}
    \label{fig:force-plaw}
\end{figure}
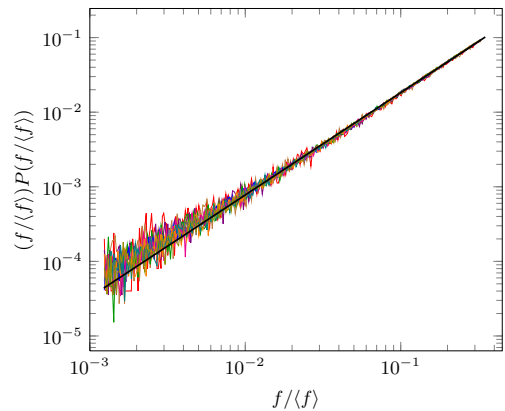

Note that the distribution seems, already at the lowest dimensions
studied here, to approach a limit distribution.
The probability density peaks around the mean value and clearly
goes to zero at zero force. Logarithmic binning of the forces
reveals a power law dependence at low force of the form
$P(f/\langle f\rangle) = c (f/\langle f\rangle)^\theta$
(see \ref{fig:force-plaw}). Again,
this behavior is consistent with the behavior of classical
jammed packings \cite{charbcorwin}. The best-fit values
of $\theta$ are given in \ref{tab:exps}
and show that the power law exponent does not
seem to depend on dimension and takes a value of $\theta=0.371\pm0.010$.

\subsection{Structure factor}

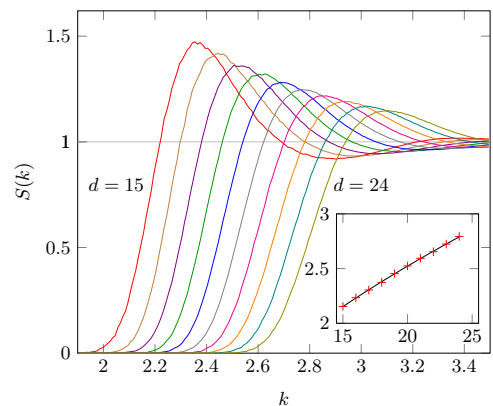
\begin{figure}
    \centering
    \scalebox{0.8}{
	\begin{tikzpicture}
	    \begin{axis}[
		xlabel={$k$}, ylabel={$S(k)$},
		domain=1.9:3.5, xmin=1.9, xmax=3.5, ymin=0.,
		y label style={at={(0.05,0.5)}},
		extra y ticks={1.},
		extra y tick style={grid=major},
	    ]
		\pgfplotsset{cycle list name=list2};
		\addplot table[x index=0,y index=1] {sk.dat};
		\addplot table[x index=0,y index=2] {sk.dat};
		\addplot table[x index=0,y index=3] {sk.dat};
		\addplot table[x index=0,y index=4] {sk.dat};
		\addplot table[x index=0,y index=5] {sk.dat};
		\addplot table[x index=0,y index=6] {sk.dat};
		\addplot table[x index=0,y index=7] {sk.dat};
		\addplot table[x index=0,y index=8] {sk.dat};
		\addplot table[x index=0,y index=9] {sk.dat};
		\addplot table[x index=0,y index=10] {sk.dat};
		\node[] at (axis cs:2.05,0.8) {$d=15$};
		\node[] at (axis cs:3.0,0.8) {$d=24$};
	    \end{axis}
	    \begin{axis}[
		width=\myinsetwidth,domain=15:24,
		xshift=0.240\textwidth, yshift=0.5cm, xmin=14.5, xmax=25.5,
		ymin=2, ymax=3,
		xticklabels = {$15$,$20$,$25$}, yticklabels = {$2$,$2.5$,$3$},
		xtick = {15,20,25}, ytick = {2,2.5,3}
	    ]
		\addplot[mark=+, only marks, red] table[x index=0,y index=1] {k0.dat};
		\addplot[] {0.483381*x^0.551549};
	    \end{axis}
	\end{tikzpicture}
    }
	\caption{
	    Spherically averaged structure factor, averaged over all jammed isostatic lattices in each dimension.
	    The inset shows the value of $k$ such that $S(k)=1/2$ as a function of dimension. The wavenumbers are
	    given in units of the inverse sphere diameter.
	}
    \label{fig:sk}
\end{figure}

We plot the spherically averaged structure factor $S(k)$, averaged
over all lattices in a given dimension, in \ref{fig:sk}. As argued in \ref{sec:thy},
the structure factor of an extreme lattice must vanish for a finite range of wave-numbers near
$k=0$, a property known as \textit{stealth} \cite{stealthy}. Classical
jammed packings, by contrast, tend to have a structure factor which approaches $0$
continuously at $k=0$ \cite{hypuni}. While
the stealth wavenumber for an extreme lattice must be greater than $\sqrt{2d/(d+1)}$
(its value for the wholly atypical lattice $A_d$), its value for
the typical jammed lattice seems to be significantly larger, and to increase
with dimension (see inset of \ref{fig:sk}). The amplitude of the oscillations about
the value $S(k)=1$ for large $k$ become smaller and smaller
with dimension, as expected by the decorrelation principle \cite{disorderwins,decorr}.

Apart from the low-wavenumber behavior,
these structure factors bare a marked resemblance to those
observed in classical jammed packing \cite{skoge,vari,benoit}. 
Those structure factor do not exhibit
stealth, but are instead heavily suppressed at small wavenumbers,
followed by a sharp rise and then oscillations about $S(k)=1$ that decay as $k\to\infty$.
The height of the first peak in $S(k)$ decreases with increasing dimension
and the wavenumber at which it appears increases.

\section{Conclusion}\label{sec:cnc}

Due in part to similarities observed between the phenomenology of structural
glasses and of the spherical p-spin glass model, there has been an
effort over the last decade to understand the glass transition and jamming in
hard-sphere systems by studying models with no spatial
extent \cite{parisi10}. In most cases, the network of interactions
between different degrees of freedom in these models are sparse graphs
or trees \cite{mf1,mf2,mf3}. However, the interactions in the spherical p-spin model
are dense: every p-tuple of spins interacts. Also, while it appears that much progress can
be made in understanding some general behavior using these models, much is lost in the process about the
distinct structural features of jamming. Features such as the power law behaviors of weak contacts
and small gaps play a major role in determining the stability of jammed states \cite{wyart,wyartsoft},
yet so far they have been missed completely in replica theory calculations \cite{charbcorwin}.
In this article, we presented a spatially nonextended model of hard spheres with
a dense interaction network that retains nearly all of the structural features of
the classical hard-sphere model. We have examined some
of the phenomena exhibited by jammed isostatic lattices, as manifested in their
densities, pair correlations, contact force distributions, and structure factors,
and the relation to similar phenomena exhibited in the classical hard-sphere model.
The lattice hard-sphere model enables numerical exploration of these phenomena
in much higher dimensions than was previously possible.
We have only begun exploring this rich set of phenomena, and there is clearly much work still to be done.

\textbf{Acknowledgements.} S. T. and {\' E}. M. were supported in part
by the Materials Research Science and Engineering Center
Program of the National Science Foundation under Grant No.
DMR-0820341 and by the Division of Mathematical Sciences
at the National Science Foundation under Award No. DMS-1211087.
This work was partially supported by a grant from the
Simons Foundation (Grant No. 231015 to Salvatore Torquato).

\bibliography{iel}

\end{document}